\documentclass[%
 reprint,
 showkeys,
 amsmath,amssymb,
aps,
rmp,
]{revtex4-1}

\usepackage{graphicx}
\usepackage{dcolumn}
\usepackage{bm}

\begin{document}


\title{Evolutionary stability and the rarity of grandmothering}

\author{Jared M. Field}
  \email{jared.field@maths.ox.ac.uk}
\affiliation{%
Wolfson Centre for Mathematical Biology, Mathematical Institute, University of Oxford, Oxford OX2 6GG,
United Kingdom \\
 }%
\altaffiliation[Also at ]{Mathematical Ecology Research Group, Department of Zoology, University of Oxford, Oxford, UK}
\author{Michael B. Bonsall}
\affiliation{Mathematical Ecology Research Group, Department of Zoology, University of Oxford, Oxford, UK}


\date{\today}

\begin{abstract}
The provision of intergenerational care, via the Grandmother Hypothesis, has been implicated in the evolution of post-fertile longevity, particularly in humans. However, if grandmothering does provide fitness benefits, a key question is why has it evolved so infrequently?  We investigate this question with a combination of life-history and evolutionary game theory. We derive simple eligibility and stability thresholds, both of which must be satisfied if intergenerational care is first to evolve and then to persist in a population. As one threshold becomes easier to fulfill, the other becomes more difficult, revealing a conflict between the two. As such, we suggest that, in fact, we should expect the evolution of grandmothering to be rare.
\end{abstract}

\maketitle

\section*{\label{sec:level1}Introduction}
Data on historical agricultural populations and modern hunter-gatherers show that these groups enjoy significant post-fertile periods \citep{alberts2013reproductive, Jones2002, gurven2007longevity, levitis2013human}. Modern medicine cannot then fully explain the life history oddity of increased longevity with reproductive inactivity. 

Taking an evolutionary approach, the Grandmother Hypothesis instead proposes that this reproductive inactivity is in fact adaptive \citep{hawkes1998grandmothering}. With the sacrifice of continued reproduction, an individual may increase their inclusive fitness by decreasing the interbirth intervals of their offspring. The care that would otherwise be put into one's own children can now be put into weaned (and increasingly independent) grandchildren, allowing their own offspring to reproduce again sooner. Otherwise put, the cost of a reduced relatedness coefficient may be outweighed by an increase in total number of grandchildren resulting from the diverted care. Several models have now shown how such a benefit could be realised \citep{KimHawkes2012,kim2014grandmothering, chan2016evolution}. In this way, a causal connection is made between the provision of intergenerational care and human post-fertile longevity. 

A valid objection to the Grandmother Hypothesis, however, is if grandmothering can result in a higher fitness, why are significant post-fertile life stages so rare? Amongst vertebrates in the wild, only humans, \emph{Globicephala macrorhynchus} (pilot whales) and \emph{Orcinus orca} (resident killer whales) have a significant proportion of individuals with such a life-history \citep{croft2015evolution}. In this paper, we present a model to investigate this objection. Our model assumes only that individuals transition through various life stages and that there is an average time to conception and gestation. In one of those stages, individuals have the option to provide care for a certain number of their grandchildren thereby allowing their own offspring to reproduce again sooner. 

By comparing inclusive fitnesses of individuals that provide intergenerational care with those that instead continue to reproduce into old age, we arrive at a necessary condition for grandmothering to be an evolutionarily stable strategy (ESS). This condition, or stability threshold, relates the number of grandchildren that care must be given to with the ratio of the length of the first two life stages. 

We then make the observation that if a grandmother is to provide care for even one set of grandchildren, their expected post-fertile stage must be sufficiently long. More precisely, for grandmothering to be adaptive, it must be the case that post-fertile life exceeds the time taken to raise a weaned child to independence. If this were not the case, grandmothers would not be able to shorten their offspring's time between births by caring for some infants themselves. In this way, we derive an eligibility threshold that tells us when grandmothering is a strategy with any possible evolutionary advantage. These eligibility and stability criteria must both be satisfied for grandmothering to evolve and then to persist.

Our analyses show that there is conflict between the stability and eligibility thresholds. As it becomes increasingly easier to meet one of them, it becomes increasingly harder to fulfill the other and \emph{vice versa}. This conflict is, at its core, a grandparent-grandoffspring conflict analogous to parent-offspring conflicts \citep{trivers1974parent}. The result of this is that there is a narrow range over which we should expect grandmothering to evolve and then to persist. In other words, we should in fact expect grandmothering to be rare. 

The rest of this paper is organised as follows: In the next section we lay out our model and assumptions. Following this, we explicitly calculate the expected inclusive fitness for the two different strategies. We then find the evolutionary stability threshold, noting that if grandmothering is to be immune to evolutionary cheating, the regular grandmothering strategy should have a higher fitness. In the proceeding section, we derive the eligibility threshold. We then use ancestral parameter values to explicitly calculate these thresholds, demonstrating the conflict between the two. Finally, we summarise our findings and suggest potential tests for the Grandmother Hypothesis.      
  
\section*{Model}

As elsewhere \citep{KimMcQueenCoxHawkes2014}, we assume that individuals transition through six possible life-history stages; unweaned, weaned, independent, fertile, post-fertile, and frail.  If we denote the age of an individual by $x$, we can write these life stages as: unweaned $x\in [0, \tau_1)$, weaned $x \in [\tau_1, \tau_2)$, independent $x\in [\tau_2, \tau_3)$, fertile $x\in [\tau_3, \tau_4)$, post-fertile $x \in [\tau_4, \tau_5)$, and frail $x \in [\tau_5, d]$, where $d$ is some maximum expected lifespan.

Once individuals reach the post-fertile period, we assume that they provide care for some of their grandchildren. We denote the number of fertile children an individual has by $\kappa$ and the number of grandchildren a post-fertile individual can care for by $\alpha$.  As infants are highly dependent on their mothers initially (for example, on their milk in the case of mammals), we further assume that intergenerational care can only be given once any given grandchild is weaned ($x > \tau_1$).   

We will eventually compare the fitness of individuals that provide grandmothering as outlined above with others that instead continue to reproduce themselves. Such evolutionary cheaters will have an older age where their post-fertile period starts. We denote this age by $\tau_{4m}$. In this case, the later life stages will be given by: fertile $x\in [\tau_{3}, \tau_{4m})$, post-fertile $x\in [\tau_{4m}, \tau_{5})$, frail $x\in [\tau_{5}, d]$. Finally, we define the average time to conception and gestation by $\beta$.

\section*{Fitness}
If it occurs that individuals with a shorter post-fertile phase achieve a higher fitness, we should expect selection to act on the shortening of this stage, reducing it further. In such a scenario, the post-fertile stage and hence the ability to grandmother should disappear. 

As the only difference in the two strategies occurs during one stage, to compare them it is sufficient to compare their inclusive fitnesses over that stage. In particular, we focus attention on the period defined by 
\begin{equation}
\tau_{4m} - \tau_{4}.
\end{equation}
In the absence of grandmothering, an individual will have to raise their own infants to $\tau_2$ (independence). As the average time to conception and gestation is $\beta$, over our period of interest an individual will be able to produce 
\begin{equation}
\frac{\tau_{4m} - \tau_{4}}{\beta + \tau_2}
\end{equation}
infants. 

Similarly, their $\kappa$ fertile children will be able to produce the same amount. Thus, the inclusive fitness over that period of an individual without grandmothering will be 
\begin{equation}
r_m = \frac{1}{2}\left(\frac{\tau_{4m} - \tau_{4}}{\beta + \tau_2}\right) + \frac{\kappa}{4} \left( \frac{\tau_{4m} - \tau_{4}}{\beta + \tau_2}\right), \label{r1}
\end{equation}
where we have added the appropriate relatedness coefficients to distinguish children and grandchildren. 

In the alternative scenario, an individual does not produce any infants themselves over our period of interest. Instead, they provide care for $\alpha$ of their grandchildren, allowing $\alpha$ of their own children to reproduce earlier than $\tau_2$ at age $\tau_1$. The remainder of their children (if there are any) will, however, have to raise their infants to $\tau_2$. Hence, the inclusive fitness of an individual that grandmothers as usual will be 
\begin{equation}
r = \frac{1}{4}\left( \alpha \left(\frac{\tau_{4m} - \tau_{4}}{\beta + \tau_1}\right) + \left(\kappa - \alpha\right)\left(\frac{\tau_{4m} - \tau_{4}}{\beta + \tau_2}\right) \right),\label{r2}
\end{equation} 
where again, the weight out the front accounts for relatedness.

\section*{Evolutionary stability threshold}
For grandmothering to be an ESS and immune to evolutionary cheating \citep{smith1982evolution}, it must be that the fitness benefits of providing intergenerational care outweigh the costs of not continuing one's own reproduction. In other words, it must be that the fitness of the regular grandmothering strategy is higher than the strategy with a reduced post-fertile stage so that
\begin{equation}
r > r_m.
\end{equation}
Using \eqref{r1} and \eqref{r2} and rearranging we find that 
\begin{equation}
\alpha > 2 \left(\frac{\beta + \tau_1}{\tau_2 - \tau_1}\right).
\end{equation}

There is hence a threshold value of number of grandchildren that must be cared for if grandmothering is to be maintained in a population. Further, this threshold depends crucially on the ratio of the time it takes to produce and wean an infant and the duration of the weaned stage. For example, for this threshold to be less than one, the average weaned stage of a species must be at least two times as long as the stage previous. Whether or not this condition is met depends on the ecology, physiology and developmental speed of a species.


 
\section*{Life history threshold}
If a grandmother is to successfully care for at least one set of grandchildren, they must live long enough. More precisely, it should be that the expected post-fertile period is longer than the weaned period so that
\begin{equation}
\tau_5 - \tau_4 > \tau_2 - \tau_1.
\end{equation}
If the post-fertile period were less than the weaned period, the grandmother would die before any of the infants they are caring for reach independence, resulting in their likely death too. In this case, a grandmothering strategy cannot provide any evolutionary advantage. 

It is important to note that this eligibility threshold is in opposition to the stability threshold. The stability threshold becomes increasingly easier to meet as the weaned period $\tau_2 - \tau_1$ increases. However, an increase in the same stage makes the eligibility threshold more difficult to achieve. While it is in the interest of the infant to have an increasingly higher $\tau_2$, grandmothers will spread more of their genes if this age is lower. In other words, there exists a grandparent-grandoffspring conflict entirely akin to parent-offspring conflicts \citep{trivers1974parent}. We suggest that this conflict goes some way towards explaining the rarity of grandmothering. 


\section*{Ancestral Parameter estimates}
In the case of ancestral humans, a previous study has estimated $\beta$ to be approximately one year \citep{KimMcQueenCoxHawkes2014}. This is found by assuming an average time to conception of half a year which is added to an average taken over the gestation times of humans, gorillas and chimpanzees. Age of weaning, $\tau_1$ is taken to be two years. This is based on the observation that in some human populations, after this age a mother's death does not increase offspring mortality \citep{sear2008keeps}. Additionally, it has been noted that chimpanzees can survive the death of their mother at this age if (a rare event) they are adopted \citep{Mace2000}. 
\begin{figure}

\caption{Minimum number of infants a grandmother must provide care for (solid line) and minimum post-fertile duration (dotted line) as a function of age of independence. Both must be met for grandmothering to persist. Here $\tau_2 = 2$ and $\beta = 1$. 
}

\includegraphics[scale=0.6]{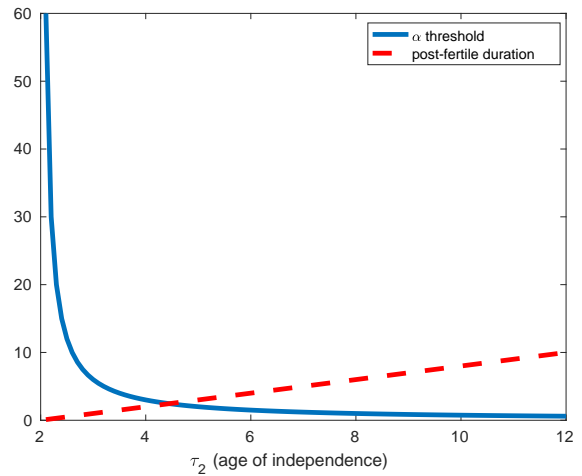}
\label{fig:Fig2}
\end{figure} 
With these values, we can explicitly calculate the stability threshold (solid line) for any value of $\tau_2$ as in Fig. \ref{fig:Fig2}. Observe that as $\tau_2$ increases, grandmothers must care for \emph{significantly} fewer infants for grandmothering to provide evolutionary benefits. The eligibility threshold (dotted line) is also plotted in Fig. \ref{fig:Fig2} for the same ancestral parameter estimates. In contrast to the stability threshold, it increases as a function of $\tau_2$.

Notice that, in Fig. \ref{fig:Fig2}, in the region to the right of $\tau_2 =8$ a grandmother must only care for one infant for grandmothering to be an ESS. However, here the post-fertile period must be in excess of six years. By contrast, to the left where the post-fertile period can be shorter, a grandmother must care for more than one. This might be fulfilled if, as the Grandmother Hypothesis suggests, the environment of our ancestors changed. Further left still however we see that the stability threshold eventually becomes biologically unrealistic. Only through an intermediate range are both thresholds biologically realisable. 

\section*{Discussion}
Intergenerational care, via the Grandmother Hypothesis, has been implicated in the evolution of post-fertile human longevity \citep{hawkes1998grandmothering, hawkes2003grandmothers, kim2012increased, KimMcQueenCoxHawkes2014}. The  extension of lifespan without an extension of fertility may be evolutionarily beneficial if, by caring for grandchildren,  the interbirth intervals of one's children are shortened. A valid objection to this hypothesis, however, is if grandmothering does allow an organism to spread more of its genes, why is it so rare? Here, we developed a simple quantitative model to investigate this objection. 

This model assumed that individuals transition through six possible life-history stages and that there is an average time to conception and of gestation. With this set-up, we noted that if a grandmother strategy is to allow the provision of enough care, the post-fertile stage should be longer than the weaned staged of their grandchildren. Indeed, it is also in the interest of children, who have twice the amount of genetic material at stake, for this to be the case. This led to a simple eligibility threshold.

We additionally asked the question, assuming grandmothering has evolved, when will it persist. Competing with individuals that continue reproducing into old age, we found that grandmothering will achieve a higher fitness only if care can be provided for a threshold number of grandchildren. This stability threshold depended on the ratio of the time it takes to produce and wean an infant and the duration of the weaned stage.

For grandmothering to evolve in the first place, and then for it to persist, both thresholds need to be met. Crucially, however, there is conflict between these two conditions. As it becomes increasingly easier to meet one, it becomes increasingly more difficult to meet the other. 

Taking ancestral parameter estimates available in the literature \citep{KimMcQueenCoxHawkes2014}, we then explicitly calculated both thresholds as functions of the age where individuals become independent. This highlighted that, for our ancestors, there was a small window of opportunity for grandmothering to evolve and persist. Our analyses have hence shown that in fact one should instead \emph{expect} grandmothering to evolve infrequently. 

This window, if the Grandmother Hypothesis is correct, was realised because of a fortuitous intersection of ecology and phylogeny. In particular, the Grandmother Hypothesis suggests that savanna-like environments, which increased during the Pliocene epoch, led our ancestors to subsist on plant foods that were manageable by older and bigger individuals but not by juveniles \citep{hawkes2005human}. This may have allowed the stability threshold to be met, particularly with economies of scale arising from grandchildren approaching independence. The eligibility threshold, if they live long enough, is also met by our closest relatives the chimpanzees \citep{cohen2004female, robson2008hominin}. However, chimpanzees that do have post-fertile periods are not the norm but the exception; overall post-reproductive representation is low \citep{levitis2013human}. Nonetheless,  it is possible that our last common ancestor also fulfilled this criterion. In this case,  the question of `why us and not them?' can be answered by ecology and in particular the stability threshold we derived. 

Unlike previous modelling on this topic, the simple thresholds of this paper all involve life history traits that can be measured. In this way, our work aims to make the evolutionary view of grandmothering testable. If grandmothering is observed and these conditions are not met, it would suggest that grandmothering is occurring for different reasons. This would then in turn cast doubt on the link between intergenerational care and post-fertile longevity. Further, the results of this paper suggest particular scenarios where we might search for non-human animals that grandmother. Additionally, once data are available, these thresholds could be used to see if the Grandmother Hypothesis can apply to other organisms (such as pilot and killer whales) that we know to have significant post-reproductive representation. 

At the heart our results is a grandparent-grandoffspring conflict that results in difficulty in fulfilling both necessary thresholds. While the literature on parent-offspring conflicts is prolific, formal work on intergenerational conflicts appears comparatively scant. In future work, it will be interesting to fully tease out the ramifications of such a conflict.



\section*{Competing interests}
We have no competing interests.
\section*{Authors' contributions}
JMF carried out the research. JMF and MBB wrote the manuscript.
\section*{Funding}
JMF is funded by the Charles Perkins Scholarship with additional financial support from UTS, Sydney.
\section*{Acknowledgments} We thank Thomas W. Scott for valuable comments and discussion.

\bibliographystyle{apsrmp4-1}
\bibliography{ref_2016_abbrv}

\begin{thebibliography}{19}%
\makeatletter
\providecommand \@ifxundefined [1]{%
 \@ifx{#1\undefined}
}%
\providecommand \@ifnum [1]{%
 \ifnum #1\expandafter \@firstoftwo
 \else \expandafter \@secondoftwo
 \fi
}%
\providecommand \@ifx [1]{%
 \ifx #1\expandafter \@firstoftwo
 \else \expandafter \@secondoftwo
 \fi
}%
\providecommand \natexlab [1]{#1}%
\providecommand \enquote  [1]{``#1''}%
\providecommand \bibnamefont  [1]{#1}%
\providecommand \bibfnamefont [1]{#1}%
\providecommand \citenamefont [1]{#1}%
\providecommand \href@noop [0]{\@secondoftwo}%
\providecommand \href [0]{\begingroup \@sanitize@url \@href}%
\providecommand \@href[1]{\@@startlink{#1}\@@href}%
\providecommand \@@href[1]{\endgroup#1\@@endlink}%
\providecommand \@sanitize@url [0]{\catcode `\\12\catcode `\$12\catcode
  `\&12\catcode `\#12\catcode `\^12\catcode `\_12\catcode `\%12\relax}%
\providecommand \@@startlink[1]{}%
\providecommand \@@endlink[0]{}%
\providecommand \url  [0]{\begingroup\@sanitize@url \@url }%
\providecommand \@url [1]{\endgroup\@href {#1}{\urlprefix }}%
\providecommand \urlprefix  [0]{URL }%
\providecommand \Eprint [0]{\href }%
\providecommand \doibase [0]{http://dx.doi.org/}%
\providecommand \selectlanguage [0]{\@gobble}%
\providecommand \bibinfo  [0]{\@secondoftwo}%
\providecommand \bibfield  [0]{\@secondoftwo}%
\providecommand \translation [1]{[#1]}%
\providecommand \BibitemOpen [0]{}%
\providecommand \bibitemStop [0]{}%
\providecommand \bibitemNoStop [0]{.\EOS\space}%
\providecommand \EOS [0]{\spacefactor3000\relax}%
\providecommand \BibitemShut  [1]{\csname bibitem#1\endcsname}%
\let\auto@bib@innerbib\@empty
\bibitem [{\citenamefont {Alberts}\ \emph {et~al.}(2013)\citenamefont
  {Alberts}, \citenamefont {Altmann}, \citenamefont {Brockman}, \citenamefont
  {Cords}, \citenamefont {Fedigan}, \citenamefont {Pusey}, \citenamefont
  {Stoinski}, \citenamefont {Strier}, \citenamefont {Morris},\ and\
  \citenamefont {Bronikowski}}]{alberts2013reproductive}%
  \BibitemOpen
  \bibfield  {author} {\bibinfo {author} {\bibnamefont {Alberts}, \bibfnamefont
  {S.~C.}}, \bibinfo {author} {\bibfnamefont {J.}~\bibnamefont {Altmann}},
  \bibinfo {author} {\bibfnamefont {D.~K.}\ \bibnamefont {Brockman}}, \bibinfo
  {author} {\bibfnamefont {M.}~\bibnamefont {Cords}}, \bibinfo {author}
  {\bibfnamefont {L.~M.}\ \bibnamefont {Fedigan}}, \bibinfo {author}
  {\bibfnamefont {A.}~\bibnamefont {Pusey}}, \bibinfo {author} {\bibfnamefont
  {T.~S.}\ \bibnamefont {Stoinski}}, \bibinfo {author} {\bibfnamefont {K.~B.}\
  \bibnamefont {Strier}}, \bibinfo {author} {\bibfnamefont {W.~F.}\
  \bibnamefont {Morris}}, \ and\ \bibinfo {author} {\bibfnamefont {A.~M.}\
  \bibnamefont {Bronikowski}}} (\bibinfo {year} {2013}),\ \href@noop {}
  {\bibfield  {journal} {\bibinfo  {journal} {Proc. Natl. Acad. Sci.}\ }\textbf
  {\bibinfo {volume} {110}}~(\bibinfo {number} {33}),\ \bibinfo {pages}
  {13440}}\BibitemShut {NoStop}%
\bibitem [{\citenamefont {Blurton~Jones}\ \emph {et~al.}(2002)\citenamefont
  {Blurton~Jones}, \citenamefont {Hawkes},\ and\ \citenamefont
  {O'Connell}}]{Jones2002}%
  \BibitemOpen
  \bibfield  {author} {\bibinfo {author} {\bibnamefont {Blurton~Jones},
  \bibfnamefont {N.~G.}}, \bibinfo {author} {\bibfnamefont {K.}~\bibnamefont
  {Hawkes}}, \ and\ \bibinfo {author} {\bibfnamefont {J.~F.}\ \bibnamefont
  {O'Connell}}} (\bibinfo {year} {2002}),\ \href@noop {} {\bibfield  {journal}
  {\bibinfo  {journal} {Am. J. Hum. Biol.}\ }\textbf {\bibinfo {volume}
  {14}}~(\bibinfo {number} {2}),\ \bibinfo {pages} {184}}\BibitemShut {NoStop}%
\bibitem [{\citenamefont {Chan}\ \emph {et~al.}(2016)\citenamefont {Chan},
  \citenamefont {Hawkes},\ and\ \citenamefont {Kim}}]{chan2016evolution}%
  \BibitemOpen
  \bibfield  {author} {\bibinfo {author} {\bibnamefont {Chan}, \bibfnamefont
  {M.~H.}}, \bibinfo {author} {\bibfnamefont {K.}~\bibnamefont {Hawkes}}, \
  and\ \bibinfo {author} {\bibfnamefont {P.~S.}\ \bibnamefont {Kim}}} (\bibinfo
  {year} {2016}),\ \href@noop {} {\bibfield  {journal} {\bibinfo  {journal} {J.
  Theor. Biol.}\ }\textbf {\bibinfo {volume} {393}},\ \bibinfo {pages}
  {145}}\BibitemShut {NoStop}%
\bibitem [{\citenamefont {Cohen}(2004)}]{cohen2004female}%
  \BibitemOpen
  \bibfield  {author} {\bibinfo {author} {\bibnamefont {Cohen}, \bibfnamefont
  {A.~A.}}} (\bibinfo {year} {2004}),\ \href@noop {} {\bibfield  {journal}
  {\bibinfo  {journal} {Biol Rev Camb Philos Soc}\ }\textbf {\bibinfo {volume}
  {79}}~(\bibinfo {number} {04}),\ \bibinfo {pages} {733}}\BibitemShut
  {NoStop}%
\bibitem [{\citenamefont {Croft}\ \emph {et~al.}(2015)\citenamefont {Croft},
  \citenamefont {Brent}, \citenamefont {Franks},\ and\ \citenamefont
  {Cant}}]{croft2015evolution}%
  \BibitemOpen
  \bibfield  {author} {\bibinfo {author} {\bibnamefont {Croft}, \bibfnamefont
  {D.~P.}}, \bibinfo {author} {\bibfnamefont {L.~J.}\ \bibnamefont {Brent}},
  \bibinfo {author} {\bibfnamefont {D.~W.}\ \bibnamefont {Franks}}, \ and\
  \bibinfo {author} {\bibfnamefont {M.~A.}\ \bibnamefont {Cant}}} (\bibinfo
  {year} {2015}),\ \href@noop {} {\bibfield  {journal} {\bibinfo  {journal}
  {Trends Ecol. Evol.}\ }\textbf {\bibinfo {volume} {30}}~(\bibinfo {number}
  {7}),\ \bibinfo {pages} {407}}\BibitemShut {NoStop}%
\bibitem [{\citenamefont {Gurven}\ and\ \citenamefont
  {Kaplan}(2007)}]{gurven2007longevity}%
  \BibitemOpen
  \bibfield  {author} {\bibinfo {author} {\bibnamefont {Gurven}, \bibfnamefont
  {M.}}, \ and\ \bibinfo {author} {\bibfnamefont {H.}~\bibnamefont {Kaplan}}}
  (\bibinfo {year} {2007}),\ \href@noop {} {\bibfield  {journal} {\bibinfo
  {journal} {Popul. Dev. Rev}\ }\textbf {\bibinfo {volume} {33}}~(\bibinfo
  {number} {2}),\ \bibinfo {pages} {321}}\BibitemShut {NoStop}%
\bibitem [{\citenamefont {Hawkes}(2003)}]{hawkes2003grandmothers}%
  \BibitemOpen
  \bibfield  {author} {\bibinfo {author} {\bibnamefont {Hawkes}, \bibfnamefont
  {K.}}} (\bibinfo {year} {2003}),\ \href@noop {} {\bibfield  {journal}
  {\bibinfo  {journal} {Am J Hum Biol}\ }\textbf {\bibinfo {volume}
  {15}}~(\bibinfo {number} {3}),\ \bibinfo {pages} {380}}\BibitemShut {NoStop}%
\bibitem [{\citenamefont {Hawkes}\ and\ \citenamefont
  {Blurton~Jones}(2005)}]{hawkes2005human}%
  \BibitemOpen
  \bibfield  {author} {\bibinfo {author} {\bibnamefont {Hawkes}, \bibfnamefont
  {K.}}, \ and\ \bibinfo {author} {\bibfnamefont {N.}~\bibnamefont
  {Blurton~Jones}}} (\bibinfo {year} {2005}),\ \href@noop {} {\bibfield
  {journal} {\bibinfo  {journal} {Grandmotherhood: the evolutionary
  significance of the second half of female life}\ }\textbf {\bibinfo {volume}
  {118}}}\BibitemShut {NoStop}%
\bibitem [{\citenamefont {Hawkes}\ \emph {et~al.}(1998)\citenamefont {Hawkes},
  \citenamefont {O’Connell}, \citenamefont {Jones}, \citenamefont {Alvarez},\
  and\ \citenamefont {Charnov}}]{hawkes1998grandmothering}%
  \BibitemOpen
  \bibfield  {author} {\bibinfo {author} {\bibnamefont {Hawkes}, \bibfnamefont
  {K.}}, \bibinfo {author} {\bibfnamefont {J.~F.}\ \bibnamefont {O’Connell}},
  \bibinfo {author} {\bibfnamefont {N.~B.}\ \bibnamefont {Jones}}, \bibinfo
  {author} {\bibfnamefont {H.}~\bibnamefont {Alvarez}}, \ and\ \bibinfo
  {author} {\bibfnamefont {E.~L.}\ \bibnamefont {Charnov}}} (\bibinfo {year}
  {1998}),\ \href@noop {} {\bibfield  {journal} {\bibinfo  {journal} {Proc.
  Natl. Acad. Sci.}\ }\textbf {\bibinfo {volume} {95}}~(\bibinfo {number}
  {3}),\ \bibinfo {pages} {1336}}\BibitemShut {NoStop}%
\bibitem [{\citenamefont {Kim}\ \emph {et~al.}(2012{\natexlab{a}})\citenamefont
  {Kim}, \citenamefont {Coxworth},\ and\ \citenamefont
  {Hawkes}}]{KimHawkes2012}%
  \BibitemOpen
  \bibfield  {author} {\bibinfo {author} {\bibnamefont {Kim}, \bibfnamefont
  {P.~S.}}, \bibinfo {author} {\bibfnamefont {J.~E.}\ \bibnamefont {Coxworth}},
  \ and\ \bibinfo {author} {\bibfnamefont {K.}~\bibnamefont {Hawkes}}}
  (\bibinfo {year} {2012}{\natexlab{a}}),\ \href@noop {} {\bibfield  {journal}
  {\bibinfo  {journal} {Proc. R. Soc. B}\ }\textbf {\bibinfo {volume}
  {279}}~(\bibinfo {number} {1749}),\ \bibinfo {pages} {4880}}\BibitemShut
  {NoStop}%
\bibitem [{\citenamefont {Kim}\ \emph {et~al.}(2012{\natexlab{b}})\citenamefont
  {Kim}, \citenamefont {Coxworth},\ and\ \citenamefont
  {Hawkes}}]{kim2012increased}%
  \BibitemOpen
  \bibfield  {author} {\bibinfo {author} {\bibnamefont {Kim}, \bibfnamefont
  {P.~S.}}, \bibinfo {author} {\bibfnamefont {J.~E.}\ \bibnamefont {Coxworth}},
  \ and\ \bibinfo {author} {\bibfnamefont {K.}~\bibnamefont {Hawkes}}}
  (\bibinfo {year} {2012}{\natexlab{b}}),\ \href@noop {} {\bibfield  {journal}
  {\bibinfo  {journal} {Proc. R. Soc. B}\ }\textbf {\bibinfo {volume}
  {279}}~(\bibinfo {number} {1749}),\ \bibinfo {pages} {4880}}\BibitemShut
  {NoStop}%
\bibitem [{\citenamefont {Kim}\ \emph {et~al.}(2014{\natexlab{a}})\citenamefont
  {Kim}, \citenamefont {McQueen}, \citenamefont {Coxworth},\ and\ \citenamefont
  {Hawkes}}]{kim2014grandmothering}%
  \BibitemOpen
  \bibfield  {author} {\bibinfo {author} {\bibnamefont {Kim}, \bibfnamefont
  {P.~S.}}, \bibinfo {author} {\bibfnamefont {J.~S.}\ \bibnamefont {McQueen}},
  \bibinfo {author} {\bibfnamefont {J.~E.}\ \bibnamefont {Coxworth}}, \ and\
  \bibinfo {author} {\bibfnamefont {K.}~\bibnamefont {Hawkes}}} (\bibinfo
  {year} {2014}{\natexlab{a}}),\ \href@noop {} {\bibfield  {journal} {\bibinfo
  {journal} {J. Theor. Biol.}\ }\textbf {\bibinfo {volume} {353}},\ \bibinfo
  {pages} {84}}\BibitemShut {NoStop}%
\bibitem [{\citenamefont {Kim}\ \emph {et~al.}(2014{\natexlab{b}})\citenamefont
  {Kim}, \citenamefont {McQueen}, \citenamefont {Coxworth},\ and\ \citenamefont
  {Hawkes}}]{KimMcQueenCoxHawkes2014}%
  \BibitemOpen
  \bibfield  {author} {\bibinfo {author} {\bibnamefont {Kim}, \bibfnamefont
  {P.~S.}}, \bibinfo {author} {\bibfnamefont {J.~S.}\ \bibnamefont {McQueen}},
  \bibinfo {author} {\bibfnamefont {J.~E.}\ \bibnamefont {Coxworth}}, \ and\
  \bibinfo {author} {\bibfnamefont {K.}~\bibnamefont {Hawkes}}} (\bibinfo
  {year} {2014}{\natexlab{b}}),\ \href@noop {} {\bibfield  {journal} {\bibinfo
  {journal} {J. Theor. Biol.}\ }\textbf {\bibinfo {volume} {353}},\ \bibinfo
  {pages} {84}}\BibitemShut {NoStop}%
\bibitem [{\citenamefont {Levitis}\ \emph {et~al.}(2013)\citenamefont
  {Levitis}, \citenamefont {Burger},\ and\ \citenamefont
  {Lackey}}]{levitis2013human}%
  \BibitemOpen
  \bibfield  {author} {\bibinfo {author} {\bibnamefont {Levitis}, \bibfnamefont
  {D.~A.}}, \bibinfo {author} {\bibfnamefont {O.}~\bibnamefont {Burger}}, \
  and\ \bibinfo {author} {\bibfnamefont {L.~B.}\ \bibnamefont {Lackey}}}
  (\bibinfo {year} {2013}),\ \href@noop {} {\bibfield  {journal} {\bibinfo
  {journal} {Evol. Anthro.}\ }\textbf {\bibinfo {volume} {22}}~(\bibinfo
  {number} {2}),\ \bibinfo {pages} {66}}\BibitemShut {NoStop}%
\bibitem [{\citenamefont {Mace}(2000)}]{Mace2000}%
  \BibitemOpen
  \bibfield  {author} {\bibinfo {author} {\bibnamefont {Mace}, \bibfnamefont
  {R.}}} (\bibinfo {year} {2000}),\ \href@noop {} {\bibfield  {journal}
  {\bibinfo  {journal} {Anim. Behav.}\ }\textbf {\bibinfo {volume}
  {59}}~(\bibinfo {number} {1}),\ \bibinfo {pages} {1}}\BibitemShut {NoStop}%
\bibitem [{\citenamefont {Maynard~Smith}(1982)}]{smith1982evolution}%
  \BibitemOpen
  \bibfield  {author} {\bibinfo {author} {\bibnamefont {Maynard~Smith},
  \bibfnamefont {J.}}} (\bibinfo {year} {1982}),\ \href@noop {} {\emph
  {\bibinfo {title} {Evolution and the Theory of Games}}}\ (\bibinfo
  {publisher} {Cambridge university press})\BibitemShut {NoStop}%
\bibitem [{\citenamefont {Robson}\ and\ \citenamefont
  {Wood}(2008)}]{robson2008hominin}%
  \BibitemOpen
  \bibfield  {author} {\bibinfo {author} {\bibnamefont {Robson}, \bibfnamefont
  {S.~L.}}, \ and\ \bibinfo {author} {\bibfnamefont {B.}~\bibnamefont {Wood}}}
  (\bibinfo {year} {2008}),\ \href@noop {} {\bibfield  {journal} {\bibinfo
  {journal} {J. Anat.}\ }\textbf {\bibinfo {volume} {212}}~(\bibinfo {number}
  {4}),\ \bibinfo {pages} {394}}\BibitemShut {NoStop}%
\bibitem [{\citenamefont {Sear}\ and\ \citenamefont
  {Mace}(2008)}]{sear2008keeps}%
  \BibitemOpen
  \bibfield  {author} {\bibinfo {author} {\bibnamefont {Sear}, \bibfnamefont
  {R.}}, \ and\ \bibinfo {author} {\bibfnamefont {R.}~\bibnamefont {Mace}}}
  (\bibinfo {year} {2008}),\ \href@noop {} {\bibfield  {journal} {\bibinfo
  {journal} {Evol. Hum. Behav.}\ }\textbf {\bibinfo {volume} {29}}~(\bibinfo
  {number} {1}),\ \bibinfo {pages} {1}}\BibitemShut {NoStop}%
\bibitem [{\citenamefont {Trivers}(1974)}]{trivers1974parent}%
  \BibitemOpen
  \bibfield  {author} {\bibinfo {author} {\bibnamefont {Trivers}, \bibfnamefont
  {R.~L.}}} (\bibinfo {year} {1974}),\ \href@noop {} {\bibfield  {journal}
  {\bibinfo  {journal} {Am. Zool.}\ }\textbf {\bibinfo {volume} {14}}~(\bibinfo
  {number} {1}),\ \bibinfo {pages} {249}}\BibitemShut {NoStop}%
\end{thebibliography}%
\end{document}